\documentclass[12pt]{article}

\usepackage{amssymb,amsmath}

\begin{document}

\begin{titlepage}

\begin{flushright}
\end{flushright}
\vskip 2.5cm

\begin{center}
{\Large \bf Single-Particle Quantum Mechanics of the Free\\
Klein-Gordon Equation with Lorentz Violation}
\end{center}

\vspace{1ex}

\begin{center}
{\large Brett Altschul\footnote{{\tt altschul@mailbox.sc.edu}}}

\vspace{5mm}
{\sl Department of Physics and Astronomy} \\
{\sl University of South Carolina} \\
{\sl Columbia, SC 29208} \\
\end{center}

\vspace{2.5ex}

\medskip

\centerline {\bf Abstract}

\bigskip

In spite of its problems with interactions,
the first-quantized Klein-Gordon equation is a satisfactory theory of free spinless particles. Moreover,
the usual theory may be extended to describe Lorentz-violating behavior, of the same types
that exist can in second-quantized scalar field theories. However, because the construction of the theory
requires a restriction to positive-energy modes, the Hilbert space inner product and the position operator
depend explicitly on the form of the Lorentz violation.

\bigskip

\end{titlepage}

\newpage

\section{Introduction}

Einsteinian relativity, encompassing both the special and general theories, elevates symmetries 
such as local Lorentz invariance to fundamental principles---which underlie both the laws of physics and
their mathematical representations. However, almost from the moment when Einstein introduced special
relativity in 1905, there have
been questions about whether the
theory's Lorentz symmetry is really exact, or whether it might just be an
extremely accurate and useful approximation. Since then, the study of apparent symmetries that
are ultimately found to be only approximately valid has become a major motif in fundamental physics.
Therefore, it is arguably even more natural now than it was in 1905 to consider whether local
Lorentz symmetry is also only approximate.

Since the 1990s, thanks to the modernization of effective field theory (EFT), it has become
relatively straightforward to lay out a test theory that allows for potential violations of
rotation invariance and Lorentz boost invariance, all in a
systematic way. The theoretical developments have been followed by a surge in experimental interest
in testing Lorentz symmetry, because the EFT approach revealed that there were many more possible
forms of Lorentz violation than had previously been appreciated; for
decades, large areas of the EFT parameter space had gone effectively unstudied. Thus far, the revived
experimental interest has not yielded any convincing positive evidence that Lorentz invariance is
not absolute, but precision Lorentz tests remain an important area of fundamental physics research.
Part of the reason for this is that the payoff if fundamental Lorentz violation is uncovered is
expected to be very high; finding Lorentz symmetry to be broken would be colossally important and would
presumably open up whole new vistas for studying the fundamental laws of the universe.

The local effective field theory for describing generalized Lorentz-violating modifications to
processes involving standard model fields is now well known~\cite{ref-kost1,ref-kost2}.
This theory, called the standard model extension (SME), is also capable of describing
forms of CPT violation involving
standard model fields that are also stable, unitary, and local, since there are close
connections between CPT violation and
Lorentz violation~\cite{ref-greenberg}.
The SME can also be expanded to cover gravitation, but the extension to metric gravity
introduces many complications that do not appear to exist in the particle sector of the SME,
which is a quantum field theory (QFT) much like the standard model---rather than a geometric theory
like general relativity (GR), which is thus far only really understood at the classical level.
The minimal SME is that subsector of the SME which is expected to be
renormalizable, as it contains only the finite number of local, Hermitian, gauge-invariant operators
which are
constructed from the standard model's known fermion and boson fields, and which are of
mass dimension four or less, making them renormalizable by power counting (so long as they
are unitary). The terms in the minimal SME action look very much like
those found in the usual standard
model action, with the key difference that the new SME operators may exhibit uncontracted Lorentz indices.
In many situations, the minimal SME is the most
natural test theory in which to analyze the results of experimental tests of Lorentz invariance and CPT.

The SME is a field theory, but a great deal may be learned about the 
physics of the minimal SME by looking at simpler physical theories with the same characteristic
Lorentz-violating kinematics.
The goal of this paper is to extend this kind of study into a new area---by looking at Lorentz violation
in a first-quantized Klein-Gordon theory. This is a timely undertaking, since there have recently been
several publications looking at solving the single-particle Klein-Gordon equation in various types of
Lorentz-violating
backgrounds~\cite{ref-bakke1,ref-vitoria1,ref-vitoria2,ref-vitoria3,ref-vitoria4,ref-ahmed1,ref-wang1,ref-zare}.
The Klein-Gordon equation has well-known problems that make it
unsuitable as a relativistic generalization of the single-particle Schr\"{o}dinger equation---at least in
the presence of interactions. However, the noninteracting
first-quantized Klein-Gordon theory is completely well
behaved, although it has a number of potentially counterintuitive features.

This paper is organized as follows.
In section~\ref{sec-finsler} we examine, by way of comparison, the use of single-particle classical
Lagrangians with Finsler structures in velocity space as descriptions of Lorentz violation. These have
proven to be useful mathematical models, but they are, at present, fundamentally incomplete---capable of
providing a full description only of free (or nearly free) particle dynamics. With this in mind, we move in
section~\ref{sec-KG} to another theory that is potentially useful, yet incomplete, because it is also
incapable of handling interactions. This is the single-particle Klein-Gordon theory that is the main focus
of the paper. In section~\ref{sec-position}, we look particularly at the structure of the position operator
in the Lorentz-violating Klein-Gordon theory. The operator has an unusual structure even in the absence of
Lorentz violation, and the broken Lorentz symmetry adds further complications. Finally, section~\ref{sec-concl}
summarizes our conclusions and avenues for further research.

\section{The Case of Finsler Geometry}

\label{sec-finsler}

The rise in interest in Finsler geometry as a formalism for describing Lorentz violation provides some
key context for the later main focus of this paper.
One important observation about Lorentz-violating theories was that explicit Lorentz violation
is not generally compatible with the general theory of relativity---or, more generally, with any theory
in which gravitational and cosmographic effects are manifestations of spacetime being a curved
pseudo-Riemannian manifold~\cite{ref-kost12}. There is not, in general, room to augment such geometrical
theories with a form
local Lorentz violation that affects particle movement, because one of the essential feature of
geometrodynamics is that test particles' spacetime trajectories are the geodesics of the manifold. Unless
stringent further conditions are met, the Lorentz violation is inconsistent with Riemann (or Riemann-Cartan)
geometry. Those stringent conditions can be satisfied if the Lorentz symmetry breaking arises
spontaneously, or when the texture of the Lorentz violation takes other special
forms~\cite{ref-bluhm6,ref-bluhm7,ref-bluhm8,ref-bluhm9}.
However, to study explicitly broken Lorentz symmetry in a nontrivial spacetime background typically
requires generalization to manifolds that have more structure than just a Riemannian metric and Cartan
spin connection.

The most obvious generalizations of this sort are to Finsler manifolds. The qualitative
connection between local Lorentz breaking and Finsler structure has been recognized for quite a long time---since
well before the development of the SME.
However, with the publication of the ``no-go'' result~\cite{ref-kost12} for explicit
violation in Riemann-Cartan spacetimes, there was naturally a growth of interest in exploring this
connection more fully.
Unfortunately, Finsler manifolds are very challenging objects to study, and the mathematical tools that
are used for studying QFT in Minkowski or Riemannian spacetimes are not generally
available in the Finsler case. Most of the fundamental fields in the standard model are spinor fields or
gauge fields, which are defined on particular bundles---more complicated than the tangent bundle---in curved
spacetimes. Satisfactory analogues of the spinor and gauge bundles on Finsler manifolds may or may not
exist, and this means that it is very difficult to make useful statements about the behavior of quantized
fields in Finsler geometries.

Even so, there was been quite a bit of interesting work done on the physics of motion in Finsler spacetimes
with nontrivial Lorentz-violating structures in momentum space. This research has primarily focused on the
motion of classical particles with Lorentz-violating free dynamics. For example,
it is possible to use a Finsler structure
to describe the motion of a particle with a nonstandard energy-momentum relation involving two
preferred background four-vectors $a$ and $e$~\cite{ref-kost3},
\begin{equation}
\label{eq-disp-ae}
(p^{\mu}-a^{\mu})(p_{\mu}-a_{\mu})-(m-e^{\mu}p_{\mu})^{2}=0;
\end{equation}
the same dynamics are produced by a classical one-particle Lagrangian with the Finsler-like
structure~\cite{ref-kost28}
\begin{equation}
\label{eq-L-ae}
L_{ae}=-\frac{m-e\cdot a}{\sqrt{1-e^{2}}}\sqrt{\dot{x}^{2}+\frac{(e\cdot\dot{x})^{2}}{1-e^{2}}}
+\left(-a+\frac{m-e\cdot a}{1-e^{2}}e\right)\cdot\dot{x}.
\end{equation}
Here $\dot{x}$ is an appropriately defined four-velocity. The dispersion relation (\ref{eq-disp-ae})
is a special case of the most general spin-independent fermionic dispersion relation in the minimal SME,
corresponding to the EFT Lagrange density
\begin{equation}
\mathcal{L}_{acef}=\bar{\psi}\left[(i\partial_{\mu})(\gamma_{\mu}+c^{\nu\mu}\gamma_{\nu}+e^{\mu}
+if^{\mu}\gamma_{5})-(m+a^{\mu}\gamma_{\mu})\right]\psi.
\end{equation}
In fact, the full classical single-particle Lagrangian $L_{acef}$ that produces the same dispersion
relation as for the on-shell field excitations of $\mathcal{L}_{acef}$ is known, but it is an extremely
complicated generalization of (\ref{eq-L-ae}). For the theory with just a $c$ tensor background, the
dispersion relation is instead
$$\left(\eta^{\mu\nu}+2c^{\mu\nu}+c^{\alpha\mu}c_{\alpha}\,\!^{\nu}\right)p_{\mu}p_{\nu}-m^{2}=
\left(p^{\mu}+c^{\mu\nu}p_{\nu}\right)\left(p_{\mu}+c_{\mu}\,\!^{\rho}p_{\rho}\right)-m^{2}=0.$$
Incorporating a $c$ term into the single-particle $L$ simply entails, at leading order, replacing the
$\eta^{\mu\nu}$ implicit in the four-vector dot products in $L_{ae}$ with
$\eta^{\mu\nu}+\frac{1}{2}c^{(\mu\nu)}$,
where $c^{(\mu\nu)}=c^{\mu\nu}+c^{\nu\mu}$ is the symmetrized version form of the background.

Even more impressively, this approach can also be used to describe the motion of particles with internal
degrees of freedom like spin. The classical particle Lagrangians~\cite{ref-kost28,ref-kost29},
\begin{equation}
L_{b}=-m\sqrt{\dot{x}^{2}}\pm\sqrt{(b\cdot\dot{x})^{2}-b^{2}\dot{x}^{2}}
\end{equation}
dictate the dynamics of a free particle whose dispersion relation has two branches,
\begin{equation}
\left(p^{2}-b^{2}-m^{2}\right)^{2}-4(b\cdot p)^{2}+4b^{2}p^{2}=0,
\end{equation}
corresponding to the energy-momentum relations for a SME fermion field with a $b$ term
\begin{equation}
\mathcal{L}_{b}=\bar{\psi}\left(i\gamma^{\mu}\partial_{\mu}-m-b^{\mu}\gamma_{5}\gamma_{\mu}\right)\psi
\end{equation}
in the
free-field action. This provides an adequate description of the center of mass motion of a spinning particle.
More elaborate constructions may be used to create classical Lagrangians for point particles with other
types of Lorentz violation~\cite{ref-colladay5,ref-kost30,ref-russell1}, including going beyond the minimal
SME to consider higher-dimensional field operators~\cite{ref-schreck2}. It is possible to draw interesting
conclusions about the
global structures of these Finsler spaces, with the modified dispersion relations living on their
tangent bundles, and there have been some first steps toward understanding vector
bundles~\cite{ref-voicu,ref-pfeifer1,ref-hohmann}.
What is missing, however, is a
dynamical picture of how a particle may pass from one branch of the dispersion relation
to another---in other words, the dynamics of a spin transition. For all the successes of the theories' modified
dispersion relations, this is a very serious deficiency. The lack of
full knowledge of how to place spin field bundles
on the Finsler spacetime means that there is, as yet, no way to study
their relativistic dynamics in the SME-Finsler framework.

\section{Klein-Gordon Theory of a Free Particle}

\label{sec-KG}

As previously noted, most of the fundamental fields in the standard model are spinor or gauge fields.
However, the standard model does have a scalar sector, in the form of the Higgs. While fields with spin
``live'' in nontrivial bundles on curved spacetimes, the scalar sector is simpler.
In particular, there are fewer possible forms
of minimal SME
Lorentz violation, since a scalar field has no intrinsic spin direction. In fact, the free dynamics of a
single complex scalar species are equivalent to those of a fermion field described by $\mathcal{L}_{acef}$
with only the $a$ and $c$
tensors nonzero; and for a real scalar field, only a nonzero $c$ is possible. The dispersion relation
for such a scalar field in Riemann-flat spacetime takes the fairly simple form (which can also be
derived from a particularly simple Finsler structure~\cite{ref-kost29})
\begin{equation}
\label{eq-disp-k}
\left(\eta^{\mu\nu}+k^{\mu\nu}\right)p_{\mu}p_{\nu}-\mu^{2}=0.
\end{equation}
This is the energy-momentum relation for the scalar QFT with Lagrange density
\begin{equation}
\label{eq-Lk}
{\cal L}=\frac{1}{2}(\partial^{\mu}
\phi)(\partial_{\mu}\phi)+\frac{1}{2}k^{\mu\nu}(\partial_{\nu}\phi)
(\partial_{\mu}\phi)-\frac{1}{2}\mu^{2}\phi^{2}.
\end{equation}
It is evident that, at leading order, $k^{\mu\nu}$ is the fermionic equivalent of $c^{(\mu\nu)}$; moreover,
the exact equivalence between boson $k$ and fermion $c$
persists, albeit with slight adjustments, to all orders. (The best way to demonstrate this
is through the use of a supersymmetry equivalence between Lorentz-violating scalar and spinor
fields~\cite{ref-berger}.) The slight differences between the dispersion relations with $k^{\mu\nu}$
and $c^{(\mu\nu)}$ arise from the fact that while $c$ may be assumed to be symmetric only at leading order, it
is manifest from $(\ref{eq-Lk})$ that the antisymmetric part of $k$ does not not actually contribute to the
action at any order.

In order to second quantize the theory defined by $(\ref{eq-Lk})$, it is often useful to rescale the
field $\phi$ and the mass parameter $\mu$ so that the Lagrange density contains no nonstandard second time
derivatives. This may be done because a term proportional to $\eta^{\mu\nu}$ may be subtracted from
$k^{\mu\nu}$ and transferred to the conventional Klein-Gordon kinetic term in $(\ref{eq-Lk})$. After
rescaling $\phi$ and $\mu$, the Lagrangian still has the general form (\ref{eq-Lk}), albeit with
different specific values for the components of the background tensor $k$. To eliminate any nonstandard
$\partial^{0}\partial^{0}$ derivatives specifically, the subtraction yields an effective $k^{\mu\nu}$
of the form $k^{\mu\nu}-k^{00}g^{\mu\nu}$. 
Although our concern here will not be with the second-quantized field theory, it will typically be
convenient to restrict attention to modified Klein-Gordon theories in which $k^{00}=0$.
Moreover, if there is only a single field in the theory [as in (\ref{eq-Lk})], it is actually
possible, via an affine redefinition of the coordinates, to eliminate all the $k^{\mu\nu}$ coefficients
from the action. This would render the Lorentz violation trivial (and, in any case, would only work in
an interacting theory for a single field sector), so we shall not follow that procedure.
Without transforming the
coordinates in this way, it is not, in general, possible to find a Lagrangian equivalent
to (\ref{eq-Lk}) in with $k^{\mu0}=0$;
However, we shall often impose this additional restriction, in order to produce tractable and
straightforwardly interpretable expressions, since separating out the positive- and negative-energy
state manifolds in a theory with time and space derivatives mixed by $k^{j0}\neq0$ may introduce
additional nontrivial complications.

Many previous studies of
the dynamics of single particles with nontrival Finsler structures in their momentum
space have treated theories that are essentially interaction free, except for the
local Lorentz violation effects associated with their Finsler metrics. However, if attention is to be
restricted to free theories, there is another kind of free theory, which (although it is little discussed
today) lies in an intermediate position between classical theories of point particles and second-quantized
field theories. This is the first-quantized scalar theory with a relativistic wave function that satisfies
the Klein-Gordon equation. Of course, the first-quantized Klein-Gordon theory has insurmountable problems
in the presence of interactions, but the free theory is perfectly well behaved, although it has some
potentially counterintuitive features. The Lorentz-violating version of the first-quantized
free Klein-Gordon theory will be the subject of the remainder of this paper. Since
a scalar field ``lives'' on the spacetime manifold itself, understanding its dynamics only requires
knowledge of the structure of the tangent bundle, which is well defined in Finsler theories. As noted, a
Lagrange density like (\ref{eq-Lk}) has itself a straightforward relationship to a Finsler structure.
The first-quantized version of the theory consequently offers a different window onto a
theory with a Finsler-like structure distorting the momentum space---a viewpoint which should be complementary
those previous analyses that have looked at the motion of single classical point masses.

For a single-particle wave function satisfying the usual relativistic Klein-Gordon equation,
\begin{equation}
\left(\partial^{\mu}\partial_{\mu}+\mu^{2}\right)\psi=0,
\end{equation}
there is a conserved vectorial current
\begin{equation}
\label{eq-KG-j}
j^{\mu}=\frac{i}{2\mu}\left[\psi^{*}\left(\partial^{\mu}\psi\right)-\left(\partial^{\mu}\psi^{*}\right)
\psi\right].
\end{equation}
At nonrelativistic energies $E\approx\mu$, the time component reduces to $j^{0}\approx\psi^{*}\psi$.
However, for wave functions that contain negative-energy Fourier components, the purported probability
density $j^{0}$ need not be positive, meaning that Born's standard probability interpretation of the
wave function $\psi$ fails completely.
In fact, because the Klein-Gordon equation is second order in time, it is
possible to specify a Cauchy initial value problem with $\dot{\psi}$ chosen arbitrarily
(subject to appropriate differentiability conditions, of course); consequently, $j^{0}$ may actually
be made arbitrarily negative in the vicinity of any given point.

The historical solution to this problem was, of course, to replace the second-order Klein-Gordon
equation with the first-order Dirac equation---thus eliminating the potentially negative time
derivative terms in $j^{0}$.
However, while some of the glaring failures of the probability current interpretation in Klein-Gordon
quantum mechanics are avoided, the single-particle Dirac theory of course has problems of its own
with the probability current (problems which typically appear in the presence of very strong interactions).

Yet there is sometimes another
way of addressing the problem of the negative probabilities that crop up
in the first-quantized Klein-Gordon theory. The
alternative approach only works as a theory of a noninteracting
Klein-Gordon particle, but we understand that even noninteracting Lorentz-violating theories may
be quite instructive; such was the case with the free
particle theories based on Finsler structure Lagrangians. The alternative solution is simply to
consider wave functions with entirely positive-frequency Fourier decompositions, and we shall explore now
how this generalizes to the Lorentz-violating theory with the dispersion relation (\ref{eq-disp-k})
instead of just $(p^{2}-\mu^{2})=0$. One immediate interesting consequence of this formulation is that
the state space of the theory is explicitly dependent on the energy-momentum relation; and thus
in the theory defined by (\ref{eq-Lk}) the space of
allowable states (and the observable operators on that state space) depend on $k$.

The modified Klein-Gordon equation with the dispersion relation (\ref{eq-disp-k}) is
\begin{equation}
\label{eq-KG-k}
\left[\left(\eta^{\mu\nu}+k^{\mu\nu}\right)\partial_{\mu}\partial_{\nu}+\mu^{2}\right]\psi=0.
\end{equation}
There is a modified current conservation equation in this theory,
with the current
\begin{equation}
j^{\mu}=-\frac{1}{2\mu}\Im\left\{\psi^{*}\left(\partial^{\mu}\psi+k^{\mu\nu}\partial_{\nu}\psi\right)\right\}
\end{equation}
generalizing (\ref{eq-KG-j}). Taking the divergence of this quantity,
\begin{equation}
\partial_{\mu}j^{\mu}=-\frac{1}{2\mu}\Im\left\{\psi^{*}\left(\eta^{\mu\nu}+k^{\mu\nu}
\right)\partial_{\mu}\partial_{\nu}\psi\right\}
\end{equation}
and applying the modified Klein-Gordon equation (\ref{eq-KG-k}), it is clear that the expression in
French brackets must be real, and hence $\partial_{\mu}j^{\mu}=0$.

If there is to be a first-quantized theory, the probability density must be the time component of the
current, $\rho\propto j^{0}$, so that $\int d^{3}x\,\rho(\vec{x}\,)$ is a conserved quantity which may be
normalized to unity. This, in turn, sets the
(unnormalized) inner product between two fields $\psi_{1}$ and $\psi_{2}$,
\begin{equation}
\label{eq-inner-prod}
\langle\psi_{2}|\psi_{1}\rangle=i\int d^{3}x\,\left\{
\psi_{2}^{*}\left[\left(\eta^{\mu0}+k^{\mu0}\right)\partial_{\mu}\psi_{1}\right]-
\left[\left(\eta^{\mu0}+k^{\mu0}\right)\partial_{\mu}\psi_{2}^{*}\right]\psi_{1}\right\}.
\end{equation}
(This expression is for a fixed time $x_{0}$, but in general, the integration may be pushed to an arbitrary
spacelike hypersufrace $\Sigma$~\cite{ref-schweber}.)
Provided $k$ is sufficiently small, there should not be problems with negativity in this theory.
Moreover (and more
precisely), in the theory in which only the space-space components $k^{jl}$ of the background are
nonvanishing, the expression for $\langle\psi_{2}|\psi_{1}\rangle$ is the same as in the
standard Lorentz-invariant theory.

Continuing henceforth in the restricted $k^{\mu0}=0$ theory, it is clear that the plane waves of positive
energy, which satisfy $i\,\Im\left\{\psi^{*}\dot\psi\right\}\geq 0$, obey the equation
\begin{equation}
\label{eq-time-schr}
i\partial^{0}\psi=\sqrt{\mu^{2}-\vec{\nabla}^{2}+k^{jl}\nabla_{j}\nabla_{l}}\,\psi.
\end{equation}
The square root operator should be understood in Fourier (three-momentum) space, such that if
\begin{equation}
\psi(x)=\int d^{3}p\,e^{i\vec{p}\cdot\vec{x}}\chi(\vec{p},x_{0}),
\end{equation}
then $\chi(\vec{p},x_{0})$ obeys
\begin{equation}
i\partial^{0}\chi(\vec{p},x_{0})=\omega(\vec{p}\,)\chi(\vec{p},x_{0})=\sqrt{\mu^{2}+\vec{p}\,^{2}-
k^{jl}p_{j}p_{l}}\,\chi(\vec{p},x_{0}).
\end{equation}
Although the equation (\ref{eq-time-schr}) is nonlocal in space, it is first order in time.
This alleviates the problem with positivity of $j^{0}$, because is no longer possible to set
$\dot{\psi}$ as an independent initial condition.

To form a wave packet out of these positive-energy plane waves, we may take a four-dimensional Fourier
expansion,
\begin{equation}
\psi(x)=\frac{\sqrt{2}}{(2\pi)^{3/2}}\int d^{4}p\,e^{-ip\cdot x}
\,\delta\!\left(p^{2}+k^{jl}p_{j}p_{l}-\mu^{2}\right)\theta(p_{0})\Psi(p).
\end{equation}
The $\delta$-function and $\theta$-function ensure that the plane wave components each obey the
Lorentz-violating Klein-Gordon equation and have positive energy; the multiplicative constants
are chosen for convenience.
To obtain a more convenient expression in terms of the Fourier-space $\Psi(p)$, we may
carry out the $p_{0}$ integration via
\begin{equation}
\delta\!\left(p_{0}^{2}-\omega(\vec{p}\,)^{2}\right)\theta(p_{0})=
\frac{1}{2\omega(\vec{p}\,)}\left[\delta\!\left(p_{0}-\omega(\vec{p}\,)\right)+
\delta\!\left(p_{0}+\omega(\vec{p}\,)\right)\right]\theta(p_{0})=
\frac{1}{2\omega(\vec{p}\,)}\delta\!\left(p_{0}-\omega(\vec{p}\,)\right)\! ,
\end{equation}
so that 
\begin{equation}
\label{eq-psi-fourier}
\psi(x)=\frac{1}{(2\pi)^{3/2}}\int\frac{d^{3}p}{\sqrt{2}\,p_{0}}e^{-ip\cdot x}\Psi(p),
\end{equation}
where $p_{0}=+\omega(\vec{p}\,)$ is now considered a function of the three-momentum $\vec{p}$. As a result,
the Fourier coefficients $\Psi(p)$ may also be written as a function $\Psi(\vec{p}\,)$ of the
three-momentum only, as there is exactly one $p_{0}>0$ value that corresponds to any given $\vec{p}$.

Returning to the inner product (\ref{eq-inner-prod})---still in the $k^{\mu0}=0$ case---it is now
possible to simplify
\begin{eqnarray}
\langle\psi_{2}|\psi_{1}\rangle & = & \int d^{3}x\left\{\psi_{2}^{*}(x)\left[\sqrt{\mu^{2}-\vec{\nabla}^{2}
+k^{jl}\nabla_{j}\nabla_{l}}\,\psi_{1}(x)\right]\right. \nonumber\\
& & \left.+\left[\sqrt{\mu^{2}-\vec{\nabla}^{2}+k^{jl}\nabla_{j}\nabla_{l}}
\,\psi_{2}^{*}(x)\right]\psi_{1}(x)\right\}
\end{eqnarray}
by inserting the Fourier transforms (\ref{eq-psi-fourier}), to yield
\begin{eqnarray}
\langle\psi_{2}|\psi_{1}\rangle & = & \frac{1}{2(2\pi)^{3}}\int\frac{d^{3}x\,d^{3}p_{2}\,d^{3}p_{1}}
{p_{0}(\vec{p}_{2})p_{0}(\vec{p}_{1})}\left\{e^{ip_{2}\cdot x}\Psi^{*}(\vec{p}_{2})
\left[\sqrt{\mu^{2}-\vec{\nabla}^{2}
+k^{jl}\nabla_{j}\nabla_{l}}\,e^{-ip_{1}\cdot x}\Psi_{1}(\vec{p}_{1})\right]\right. \nonumber\\
&  & \left.+\left[\sqrt{\mu^{2}-\vec{\nabla}^{2}+k^{jl}\nabla_{j}\nabla_{l}}
\,e^{ip_{2}\cdot x}\Psi_{2}^{*}(\vec{p}_{2})\right]e^{-ip_{1}\cdot x}\Psi_{1}(\vec{p}_{1})\right\}.
\end{eqnarray}
Collecting common terms and picking out the $\delta$-function
$(2\pi)^{-3}\!\int d^{x}\,e^{i(p_{2}-p_{1})\cdot x}$, we finally reach
\begin{equation}
\label{eq-prod-fourier}
\langle\psi_{2}|\psi_{1}\rangle=\int\frac{d^{3}p}{p_{0}}\,\Psi_{2}^{*}(\vec{p}\,)\Psi_{1}(\vec{p}\,).
\end{equation}
In this form, no Lorentz-violation coefficient $k$ appears; however, the Lorentz violation is still
present in a fundamental way.
The Lorentz-violating dispersion relation $p_{0}=+\omega(\vec{p}\,)$ enters as a weight in the integration.
So the inner product---as, for example, expanded in plane waves (\ref{eq-prod-fourier})---depends on the
character of the Lorentz violation!

Plane waves with position-independent amplitudes are, of course, not strictly normalizable. However,
we may adopt a standard $\delta$-function normalization for these eigenstates of the rigged Hilbert space.
With a continuum normalization
\begin{equation}
\langle\psi_{\vec{p}_{2}}|\psi_{\vec{p}_{1}}\rangle=p_{0}\delta^{3}\!\left(\vec{p}_{1}-\vec{p}_{2}\right),
\end{equation}
the plane waves should be
\begin{equation}
\psi_{\vec{p}}=\frac{e^{-ip\cdot x}}{\sqrt{2(2\pi)^{3}}}
\end{equation}
up to an overall phase. However, this means that the closure relation
\begin{equation}
\label{eq-closure}
\sum_{\vec{p}}|\psi_{\vec{p}}\,(x_{1})
\rangle\langle\psi_{\vec{p}}\,(x_{2})|=\frac{1}{2(2\pi)^{3}}\int\frac{d^{3}p}{p_{0}}\,
e^{-ip\cdot(x_{1}-x_{2})}
\end{equation}
is not a $\delta$-function---even when the times are equal, $x_{1}^{0}=x_{2}^{0}$. [The
momentum sum on
the left-hand side of the expression (\ref{eq-closure}) is schematic, but the expression on the right-hand
side is unambiguous. The form is dictated by requiring that when sandwiched between
an additional bra on the left and ket on the right, the resulting inner products take
the form (\ref{eq-prod-fourier}).]

Note, however, that the principal peculiarity of
(\ref{eq-closure})---that the closure relation does not yield a $\delta$-function---is not
a consequence of the Lorentz violation, but of the fact that the positive-energy modes of the Klein-Gordon
operator do not span the full space of square-integrable spacetime functions. There is the same discrepancy
in the Lorentz-symmetric theory. However, the positive-energy modes do span a Hilbert space
$\mathcal{H}_{+}$ of free particle
modes, in which the plane waves are manifestly eigenstates of the three-momentum operator.

\section{The Position Operator}

\label{sec-position}

Where the operator algebra is really modified is with the introduction of the position operator.
One of the strangest properties of having a theory with these atypical position operators is that
spatially localized wave functions are not described by $\delta$-functions [a parallel to the
lack of a $\delta$-function in (\ref{eq-closure})]. Instead, a different,
more general characterization of spatially localized states is required.

Because
the inner product in the momentum representation (\ref{eq-closure}) contains the nontrivial weighting
factor $1/p_{0}$, the operator $i\vec{\nabla}_{\vec{p}}$ is not self-adjoint and so cannot be the
representation of the physical position observable $\vec{x}\,^{{\rm op}}$,
\begin{eqnarray}
\langle\psi_{2}|x_{j}\psi_{1}\rangle & = & i\int\frac{d^{3}p}{p_{0}}\Psi_{2}^{*}(\vec{p}\,)
\frac{\partial}{\partial p_{j}}\Psi_{1}(\vec{p}\,) \\
& = & \int\frac{d^{3}p}{p_{0}}\left[\left(-i\frac{\partial}{\partial p_{j}}+i\frac{p_{j}-k_{jl}p^{l}}
{\mu^{2}+\vec{p}\,^{2}-k^{lm}p_{l}p_{m}}
\right)\Psi_{2}^{*}(\vec{p}\,)\right]\Psi_{1}(\vec{p}\,) \\
& \neq & \langle x_{j}\psi_{2}|\psi_{1}\rangle.
\end{eqnarray}
Instead, the self-adjoint part of $\vec{x}$ should be defined to be the position operator
(recall that we are still working in the restricted theory in which $k^{\mu0}=0$),
\begin{equation}
\label{eq-xop}
x^{{\rm op}}_{j}=i\frac{\partial}{\partial p_{j}}-\frac{i}{2}\frac{p_{j}-k_{jl}p^{l}}
{\mu^{2}+\vec{p}\,^{2}-k^{lm}p_{l}p_{m}}.
\end{equation}
The form of the position operator depends explicitly on the Lorentz-violating background!

This is obviously a generalization of the result from the Lorentz-invariant ($k^{\mu\nu}=0$)
first-quantized Klien-Gordon theory,
\begin{equation}
\label{eq-xop-noLV}
\vec{x}\,^{{\rm op}}=i\vec{\nabla}_{\vec{p}}-\frac{i}{2}\frac{\vec{p}}{\mu^{2}+\vec{p}\,^{2}}.
\end{equation}
The understanding that this was the proper quantum-mechanical position operator for the
single-particle Klein-Gordon theory
originally arose out a broader study of ``center-of-mass'' operators in relativistic
quantum mechanics~\cite{ref-pryce,ref-moller1,ref-newton}. Interest in this problem was, unsurprisingly, more
focused on the more complicated and physically important first-quantized Dirac theory. In that case, careful
analyses led to the identification of the
{\em Zitterbewegung}-free ``mean'' position operator---which is usually
derived using a Foldy-Wouthuysen transformation~\cite{ref-foldy,ref-ellis2}---by different means.
(The role of {\em Zitterbewegung} and how to separate it from the mean velocity
in the Lorentz-violating Dirac theory
have previously previously been discussed in Ref.~\cite{ref-altschul4}.)

In the single-particle Dirac theory, the {\em Zitterbewegung} (and the procedure for
eliminating it to extract the mean position operator) are also inseparable from the
theory's energy-momentum
relation. The high-frequency oscillations that are part of the dynamics 
(and other related quirks, such as the components of the formal velocity operator having only $\pm 1$
eigenvalues, and the velocity components not commuting with one-another) appear because of interference
between positive- and negative-frequency components in a well-localized wave function. To free the
physical position operator from the {\em Zitterbewegung} will thus necessarily requires deriving a
new mean operator that depends on the dispersion relation---similar to the way that $\vec{x}\,^{{\rm op}}$
invokes $\omega(\vec{p}\,)$ and its gradient, whereas the naive operator $i\vec{\nabla}_{\vec{p}}$
does not.

In the Klein-Gordon theory, the position operator may be characterized axiomatically~\cite{ref-schweber}
(although there are much more recent
criticisms of that approach~\cite{ref-zavialov}---related, in part, to the lack of connection to a satisfactory
interacting theory),
and one of the conditions that characterizes the spatially nonlocal position operator
(\ref{eq-xop-noLV}) is that physical states localized at different spatial points must be orthogonal, in spite
of the fact that such localized wave functions are not $\delta$-functions with point-like support.
We shall now look at the Lorentz-violating generalization of this
spatial orthogonality characterization.

A necessary condition for the theory to be interpretable
is that two states localized at different positions at a common
time $x_{0}=0$ be orthogonal. Assuming that appropriately localized states do exist, the states centered on
different points must be spatial translates of one-another. Starting with a state $|\psi_{\vec{x}_{1}}\rangle$
with momentum-space representation (\ref{eq-psi-fourier}) $\Psi_{\vec{x}_{1}}(\vec{p}\,)$
the $\vec{p}$-space representation of a translation by $\vec{x}_{2}-\vec{x}_{1}$ is simply
$e^{-i\vec{p}\cdot(\vec{x}_{2}-\vec{x}_{1})}$, so
$\Psi_{\vec{x}_{2}}(\vec{p}\,)=e^{-i\vec{p}\cdot(\vec{x}_{2}-\vec{x}_{1})}
\Psi_{\vec{x}_{1}}(\vec{p}\,)$, and the overlap of the two states is
\begin{equation}
\delta^{3}(\vec{x}_{2}-\vec{x}_{1})=\langle\psi_{\vec{x}_{2}}|\psi_{\vec{x}_{1}}\rangle=
\int\frac{d^{3}p}{p_{0}}e^{i\vec{p}\cdot(\vec{x}_{2}-\vec{x}_{1})}\Psi_{\vec{x}_{2}}^{*}(\vec{p}\,)
\Psi_{\vec{x}_{1}}(\vec{p}\,).
\end{equation}
This is indeed a $\delta$-function if $\left|\Psi_{\vec{x}_{1}}(\vec{p}\,)\right|^{2}=(2\pi)^{-3}\,p_{0}$.

This fixes the magnitude of the momentum-space wave function of a spatially localized state. To get
the phase, we must apply the eigenvalue condition for $\vec{x}\,^{{\rm op}}$ directly. [In the
Lorentz-invariant theory, it is possible to obtain the full $\Psi_{\vec{x}_{1}}(\vec{p}\,)$ in a
different way, by imposing physical $O(3)$ invariance around ${\vec{x}_{1}}$. However, that
obviously does not apply to the theory with the Lorentz-violating $k$ term.] Setting
$\Psi_{\vec{x}}(\vec{p}\,)=(2\pi)^{-3/2}\sqrt{p_{0}(\vec{p}\,)}\,e^{i\alpha(\vec{x},\vec{p}\,)}$, the
eigenvalue equation
\begin{equation}
\left(i\frac{\partial}{\partial p_{j}}-\frac{i}{2}\frac{p_{j}-k_{jl}p^{l}}
{\mu^{2}+\vec{p}\,^{2}-k^{lm}p_{l}p_{m}}\right)\sqrt{p_{0}(\vec{p}\,)}\,e^{i\alpha(\vec{x},\vec{p}\,)}=
x_{j}\sqrt{p_{0}(\vec{p}\,)}\,e^{i\alpha(\vec{x},\vec{p}\,)},
\end{equation}
implies simply that $\alpha(\vec{x},\vec{p}\,)=-i\vec{p}\cdot\vec{x}$ (plus an irrelevant constant),
since the second term in
$x_{j}^{{\rm op}}$ arose as precisely $-\frac{1}{2}$ times what was needed to cancel
$i\,\partial/\partial p_{j}$ acting on $1/p_{0}$. So the Fourier representation of the localized
spatial state is
\begin{equation}
\Psi_{\vec{x}}(\vec{p}\,)=(2\pi)^{-3/2}\left(\mu^{2}+\vec{p}\,^{2}-k^{jl}p_{j}p_{l}\right)^{1/4}
e^{-i\vec{p}\cdot\vec{x}},
\end{equation}
and the amplitude for finding a particle in the state $|\psi\rangle$ at a particular position $\vec{x}$ is
\begin{equation}
\langle\psi_{\vec{x}}|\psi\rangle=\frac{1}{(2\pi)^{3/2}}\int d^{3}p\,\frac{e^{-i\vec{p}\cdot\vec{x}}}
{\left(\mu^{2}+\vec{p}\,^{2}-k^{jl}p_{j}p_{l}\right)^{1/4}}\Psi(\vec{p}\,).
\end{equation}

The wave function in configuration space is, as previously noted, not a $\delta$-function. It does not
need to be to be completely localized, since the probability density is not $|\psi(x)|^{2}$. Instead, for
a wave function localized at the origin (still on the time slice $x_{0}=0$), the formula is
\begin{equation}
\psi_{\vec{0}}(x_{0}=0,\vec{x}\,)=\frac{1}{\sqrt{2}(2\pi)^{3/2}}\int\frac{d^{3}p}{p_{0}}\sqrt{p_{0}}
e^{i\vec{p}\cdot\vec{x}}.
\end{equation}
A very similar integral to this one (differing only by the power of $p_{0}$ in the integrand) was
previously evaluated in Ref.~\cite{ref-altschul31}, in the context of
finding the Yukawa potential mediated by a scalar field with $k$-type Lorentz violation. The
integrals may be related to their Lorentz-violating ($k^{jl}=0$) versions using a linear change of
coordinates. (This is actually closely related to the change of coordinates, discussed in
section~\ref{sec-KG} that could be used to remove $k$ from the field theory Lagrangian.)
Define a matrix $K$
with $K_{jl}=\delta_{jl}-k_{jl}$; $K$ is symmetric and, presuming the Lorentz violation is small,
also positive definite. By making an orthogonal rotation of the integration variables $\vec{p}$,
we may diagonalize it, and in the rotated coordinates, the
Fourier transform becomes
\begin{equation}
\psi_{\vec{0}}(0,\vec{x}\,)=\frac{1}{\sqrt{2}(2\pi)^{3/2}}\int d^{3}p\,
\frac{e^{i\vec{p}\cdot\vec{x}}}{\left(\mu^{2}+K_{11}p_{1}^{2}+K_{22}p_{2}^{2}+K_{33}p_{3}^{2}\right)^{1/4}}.
\end{equation}

Another linear transformation will suffice to bring the integral into a known form,
rescaling $\bar{p}_{j}=\sqrt{K_{jj}}\,p_{j}$. (In this paragraph, there is no implied sum over
the repeated index $j$, although the sum remains for other Roman indices.) The exponential
in the Fourier transform becomes 
$e^{i\bar{p}_{l}\bar{x}_{l}}$, with $\bar{x}_{j}=x_{j}/\sqrt{K_{jj}}$ (no sum). There is
also a nontrivial change in the integration measure,
\begin{equation}
d^{3}p=\frac{d^{3}\bar{p}}{\sqrt{\det K}},
\end{equation}
bringing the wave function into the form
\begin{equation}
\psi_{\vec{0}}(0,\vec{x}\,)=\frac{1}{\sqrt{2(\det K)}(2\pi)^{3/2}}
\int d^{3}\bar{p}\,\frac{e^{i\bar{p}_{l}\bar{x}_{l}}}{\left(\bar{p}_{l}\bar{p}_{l}+\mu^{2}
\right)^{1/4}}.
\end{equation}

From this point, the calculation mirrors the one in the Lorentz-invariant theory. The Fourier
transform may be expressed in terms of a Hankel function of fractional order,
\begin{equation}
\label{eq-psi0}
\psi_{\vec{0}}(0,\vec{x}\,)=\frac{\mathcal{N}}{\sqrt{\det K}}\left(\frac{\mu}{\bar{r}}\right)^{5/4}
H^{(1)}_{5/4}(i\mu\bar{r}),
\end{equation}
where the spatial dependence is contained in 
\begin{eqnarray}
\label{eq-xbar}
\bar{r}=\sqrt{\bar{x}_{j}\bar{x}_{j}}=\sqrt{\left(K^{-1}\right)_{jl}x_{j}x_{l}}
\approx|\vec{x}|+\frac{k_{jl}x_{j}x_{l}}{2|\vec{x}|}
\end{eqnarray}
(with once again a sum over $j=1,2,3$), and $\mathcal{N}$ is an overall constant.
The approximate form in (\ref{eq-xbar}) makes use of the fact that $(K^{-1})_{jl}\approx \delta_{jl}+k_{jl}$
at leading order
in $k$, in which case $\sqrt{\det K}\approx 1-k_{jj}/2$ also. For a wave function localized
at a different point $\vec{x}_{0}$, the wave function $\psi_{\vec{x}_{0}}(0,\vec{x}\,)$ still takes the
form (\ref{eq-psi0}) but with the natural replacement
\begin{equation}
\bar{r}=\sqrt{\left(K^{-1}\right)_{jl}\left(x-x_{0}\right)_{j}\left(x-x_{0}\right)_{k}}\,.
\end{equation}

Because the argument of the Hankel function of the first kind is purely imaginary, the wave spatial
wave function falls off exponentially at large distances, with a characteristic spatial extent $\sim 1/\mu$.
However, it cannot be normalized, since it
diverges as $\bar{r}^{-5/2}$ as $\bar{r}\rightarrow0$. (The power law prefactor and the Hankel function
each separately go as $\bar{r}^{-5/4}$ at small $\bar{r}$.) Qualitatively, the problems with normalizability
are quite similar to what is normally encountered with $\delta$-function-localized states; there are no
issues at long distances, but the divergence in the immediate vicinity of the point of localization makes
the states not square integrable.  The reason that $\psi_{\vec{0}}(0,\vec{x}\,)$ cannot be an
actual $\delta$-function is essentially that the state has to be built entirely out of positive-energy
solutions of the Klein-Gordon equation, and the corresponding Hilbert space $\mathcal{H}_{+}$ is not
rich enough to represent $\delta$-functions. (Something similar may be familiar from the first-quantized Dirac
theory, where it is also impossible to construct very
narrowly localized wave functions without including
negative-energy states.)

Concluding our analysis of the position operator and its eigenstates, we may look at the commutators of
the position projections $x_{j}^{{\rm op}}$ with other important operators. It is apparent from inspection
that $\vec{x}\,^{{\rm op}}$ forms the usual canonical commutation relations with the momenta $\vec{p}$,
\begin{equation}
[x_{j}^{{\rm op}},p_{l}]=i\delta_{jk}.
\end{equation}
It is also straightforward that
\begin{equation}
[x_{j}^{{\rm op}},x_{l}^{{\rm op}}]=0,
\end{equation}
since the second term on the right-hand side of (\ref{eq-xop}) is already proportional to
$\partial/\partial p_{j}$ acting on $1/p_{0}$. According to the Schr\"{o}dinger equation
(\ref{eq-time-schr}), the momentum is still a constant of the motion---no surprise for a free theory.
However, the commutator of the Hamiltonian and the position operator is nontrivial,
\begin{equation}
\dot{x}_{j}^{{\rm op}}=i[H,x_{j}^{{\rm op}}]=i[p_{0},x_{j}^{{\rm op}}]=
\frac{p_{j}-k_{jl}p^{l}}{\mu^{2}+\vec{p}\,^{2}-k^{lm}p_{l}p_{m}}.
\end{equation}
This velocity has the same form as that found for a Lorentz-violating fermion in a theory with $c$
terms~\cite{ref-altschul4}.
Moreover, it is also clearly identical to the group velocity $\vec{v}_{g}=\nabla_{\vec{p}}\,p_{0}$.

\section{Conclusions and Outlook}

\label{sec-concl}

It is evident that the first-quantized Klein-Gordon theory with Lorentz violation has a number of peculiar
features. Some of these are present even in the Lorentz-invariant theory. The
necessity of excluding negative-energy states from the theory leads to explicit dependences of many of
the theory's structures---including the inner product in momentum space, the closure relation,
and the position operator---on the energy $p_{0}$. With the inclusion of Lorentz violation, the Hamiltonian
for the theory is modified, and the Lorentz-violating structure appears in all those energy-dependent
structures. This provides a new kind of Lorentz-violating test theory that may be useful for understanding
the evolution of spinless particle states.

The theory discussed in this paper sits in an intermediate position between purely classical theories
and interacting quantum field theories. Unlike the
point-particle Lagrangian theories discussed in section~\ref{sec-finsler},
the first-quantized theory is described by a Hilbert space of states and an algebra of operators acting on
those states. However,
like theories with classical Lagrangians that give only the Lorentz-violating dispersion relations
for particle excitations, this Lorentz-violating Klein-Gordon theory is limited to treating only freely
propagating excitations. Yet knowledge of how freely-propagating particles with Lorentz-violating
dispersion relations behave can, on its own, be very powerful. For example, changes to reaction kinematics
are often the most important determiners of how Lorentz violation affects particle scattering and decay
processes~\cite{ref-kost13,ref-altschul2}.
Up to now, there has
also been relatively little examination~\cite{ref-altschul4} of how wave packet behavior may be affected
by the SME's anisotropic kinetic energy terms, and the Klein-Gordon theory, which has a complete probabilistic
interpretation, provides a new arena for such studies.

Other further examinations of Lorentz-violating Klein-Gordon solutions are also possible. 
The relationship between the wave equation (\ref{eq-KG-k}) and a Finsler spacetime
structure may be made more precise. Moreover,
there are remaining question about the Lorentz-violating Klein-Gordon theory with nonstandard time
derivatives, entering via nonzero $k^{j0}$ terms. Such terms are odd under time reversal and will, at a
minimum, affect the separation of positive- and negative-energy modes.

More generally,
when considering the possibility of a major change in how physics is understood to operate---and a
fundamental violation of Lorentz symmetry would certainly constitute such a basic change---it may be worthwhile
to understand the impact of the change from multiple viewpoints.
The theories that we use to describe physics---both putatively fundamental theories
as well as effective ones---may
generally be described mathematically in more than one way, meaning via more than just a single formalism.
When generalizing such theories,
it is not necessarily the case that the modified theories will be describable using all the formalisms that
worked for the original theories. That makes it worthwhile to study new physics using more than one
mathematical approach, and---like the application of Finsler Lagrangians for classical point particles---the
first-quantized Klien-Gordon formalism provides a potentially useful alternative viewpoint for understanding
Lorentz violation among bosons.

\end{document}